\documentclass[8pt,a4paper,twocolumn]{article}

\usepackage[utf8]{inputenc}
\usepackage[T1]{fontenc}
\usepackage{microtype}

\usepackage[top=2.5cm, bottom=2.5cm, left=2.5cm, right=2.5cm]{geometry}

\usepackage{parskip}

\usepackage{authblk}

\usepackage[colorlinks=true,
            linkcolor=blue!60!black,
            citecolor=blue!60!black,
            urlcolor=blue!60!black]{hyperref}

\usepackage[
  backend=biber,
  style=authoryear,
  natbib=true,
  maxbibnames=3,
  minbibnames=3
]{biblatex}
\usepackage{booktabs}
\usepackage{tabularx}
\usepackage{array}
\usepackage{graphicx}

\usepackage{listings}
\usepackage{xcolor}
\lstdefinestyle{jsstyle}{
  backgroundcolor=\color{gray!8},
  basicstyle=\ttfamily\footnotesize,
  keywordstyle=\color{blue!70!black}\bfseries,
  stringstyle=\color{orange!80!black},
  commentstyle=\color{gray}\itshape,
  numberstyle=\tiny\color{gray},
  numbers=left,
  numbersep=6pt,
  frame=single,
  framerule=0.4pt,
  rulecolor=\color{gray!40},
  breaklines=true,
  captionpos=b,
  language=JavaScript
}
\lstdefinestyle{shellstyle}{
  backgroundcolor=\color{gray!8},
  basicstyle=\ttfamily\footnotesize,
  frame=single,
  framerule=0.4pt,
  rulecolor=\color{gray!40},
  breaklines=true,
  captionpos=b
}

\usepackage{enumitem}
\setlist[itemize]{itemsep=2pt, topsep=4pt}

\usepackage{amsmath}

\usepackage{multicol}

\newcommand{\sce}{\textit{Survey Footprint Explorer}}
\newcommand{\moc}{\textsc{moc}}

\title{\textbf{Survey Footprint Explorer:} A Browser-Based Interactive Tool for
  Visualizing and Cross-Matching Astronomical Survey Footprints}

\author[1,2,3]{Syeda~Lammim~Ahad \footnote{slahad@uwaterloo.ca}}
\author[4,5]{Ruslan~Brilenkov}
\author[1,2]{James~E.~Taylor}

\affil[1]{Waterloo Centre for Astrophysics, University of Waterloo, Waterloo, ON N2L 3G1, Canada}
\affil[2]{Department of Physics and Astronomy, University of Waterloo, Waterloo, ON N2L 3G1, Canada}
\affil[3]{Center for Astronomy, Space Science and Astrophysics, Independent University, Bangladesh, Dhaka 1229, Bangladesh}
\affil[4]{Kapteyn Astronomical Institute, University of Groningen, Landleven 12, 9747~AD Groningen, The Netherlands}
\affil[5]{Definity Insurance Company, 111 Westmount Road South, Waterloo, ON N2J 4S4, Canada}

\date{\small{Version 2.5.0 --- \today}}

\begin{document}

\twocolumn[%
  \maketitle

% =============================================================================


% =============================================================================
    \begin{abstract}
% =============================================================================

    We present the \sce{} (v2.5.0) a browser-based interactive tool for visualising and comparing the sky footprints of major astronomical imaging surveys. The tool is implemented entirely in client-side JavaScript and requires no server infrastructure, making it immediately accessible from any modern web browser. Thirteen survey footprints are currently included: Euclid~DR1, LSST Wide-Fast-Deep, the Nancy Grace Roman HLWAS and HLTDS (full and deep tiers), DESI Legacy Imaging Survey~DR9, the Dark Energy Survey (DES), the Subaru Hyper Suprime-Cam survey (HSC), the Kilo-Degree Survey (KiDS), the Ultraviolet Near-Infrared Optical Northern Survey (UNIONS), the eROSITA All-Sky Survey (eRASS1), and the Atacama Cosmology Telescope Legacy (ACT) survey spanning wavelengths from X-ray to near-infrared and covering footprints from 7.7~deg$^{2}$ to 21\,524.4~deg$^{2}$. Survey footprints are encoded as Multi-Order Coverage (\moc{}) maps and rendered via two complementary views: an interactive globe powered by Aladin Lite~v2, and a full-sky equirectangular projection. All \moc{} intersection calculations, including multi-survey overlap area computation and per-source membership testing, are performed client-side. Users may upload source catalogues in CSV or TSV format and download an augmented version with boolean survey membership columns appended. The link to access the tool is provided at the end of the Summary section.

  \end{abstract}
  \vspace{1em}
]

%\begin{multicols}{2}
% =============================================================================
\section{Introduction}
\label{sec:intro}
% =============================================================================

Astronomy is entering an era of unprecedented large-scale and deep imaging surveys. Missions and programs such as the \textit{Euclid} space telescope \citep{euclid2025}, the Vera C.\ Rubin Observatory's Legacy Survey of Space and Time (LSST; \citealt{ivezic2019}), the Nancy Grace Roman Space Telescope High Latitude Wide Area Survey (Roman HLWAS; \citealt{akeson2019}), and the DESI Legacy Imaging Survey \citep{dey2019} together will image billions of galaxies across a significant fraction of the extragalactic sky. The scientific returns of these programs are greatly amplified by combining data across surveys -- whether for multi-wavelength characterisation, weak lensing tomography, calibration of photometric redshifts, or selection of spectroscopic targets. Any such cross-survey analysis requires a clear understanding of which regions of sky are covered by each survey, where footprints overlap, and which sources in a given catalogue fall within a given coverage area.

Survey footprints are most naturally represented using Multi-Order Coverage (\moc{}) maps \citep{fernique2014, fernique2022}, an International Virtual Observatory Alliance (IVOA) standard based on the {HEALPix} spherical pixelisation scheme \citep{gorski2005}. \moc{} maps encode sky regions as sets of HEALPix cells at adaptively chosen resolutions, enabling compact storage and fast set-theoretic operations (e.g., union, intersection, difference) on arbitrary sky regions through simple integer arithmetic. Python libraries such as \texttt{mocpy} \citep{mocpy} and \texttt{healpy} \citep{zonca2019} provide full programmatic access to \moc{} generation and set operations, but they operate on individual files and do not offer a unified environment for loading, visualising, and comparing the footprints of multiple surveys side by side.

Existing interactive tools for footprint visualisation include Aladin Desktop \citep{boch2014}, a powerful Java application, and the CDS Aladin Lite browser widget, which renders \moc{} overlays on an interactive sky globe. However, these tools do not natively support quantitative multi-survey intersection calculations, catalogue upload with per-survey membership testing, or export of an augmented catalogue to disk. \textsc{topcat} \citep{taylor2005} can handle catalogue cross-matching but requires a local installation and does not directly visualise \moc{} footprints. No lightweight, zero-install, browser-based tool currently combines all of: (i)~multi-survey footprint visualisation in two complementary projections, (ii)~client-side \moc{} intersection area computation, and (iii)~per-source catalogue membership determination with downloadable augmented output. 

Here we present the \sce{} (version 2.4.0), a static web application that addresses this gap. The tool runs entirely in the browser with no server-side computation and no software installation, relying on a WebAssembly-compiled Rust \moc{} engine for all coverage calculations. It provides interactive access to the footprints of ten major surveys spanning X-ray to near-infrared wavelengths and supports upload of custom \moc{} files and source catalogues. The remainder of this paper is organised as follows. Section~\ref{sec:description} describes the user-facing features of the tool. Section~\ref{sec:surveys} details the survey footprints currently included and their \moc{} encoding. Section~\ref{sec:implementation} describes the technical implementation. Section~\ref{sec:usecases} presents three illustrative use cases. Section~\ref{sec:conclusion} summarises and outlines future work. 

% =============================================================================
\section{Software Description}
\label{sec:description}
% =============================================================================

\subsection{Overview}

The \sce{} is a single-page web application with a two-panel layout: a fixed sidebar containing all controls, and a full-height map panel housing the sky view. The application state, including survey selection, display preferences, and any uploaded catalogue, is optionally persisted to browser \texttt{localStorage} so that the user's session is restored on return visits. Each user's application state is stored only in their browser.

\subsection{Survey Selection and Layering}

Surveys are selected via a dropdown panel containing labelled checkboxes, one per survey. Multiple surveys may be selected simultaneously; their footprint overlays are rendered in the order they appear in the list, with the survey at the top of the list drawn on top. The user may drag and reorder surveys within the list to control layering. A ``Select all'' option is available for rapid global comparison. The sidebar Coverage panel displays the number of currently selected surveys and, once computed, the intersection area of all selected footprints in square degrees (see Section~\ref{sec:wasm} for the computation method).

\begin{figure}
\centering
\includegraphics[width=0.7\linewidth]{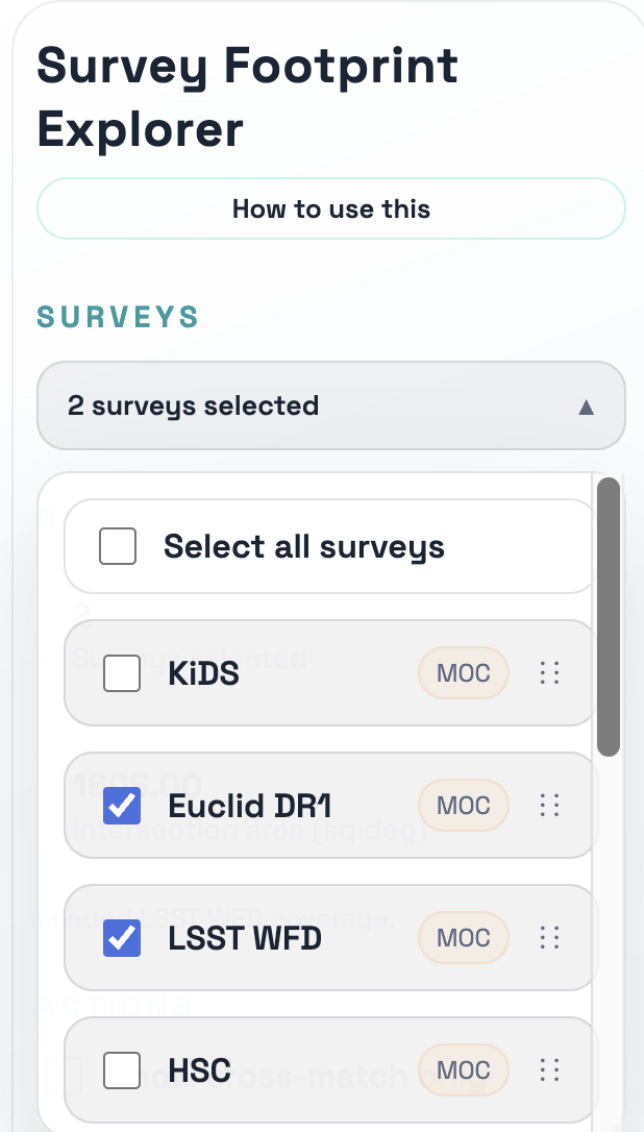}
\caption{Survey selection.}
\label{fig:survey_selection}
\end{figure}

\subsection{Map Views}
\label{sec:views}

\subsubsection{Aladin Globe View}

The primary view embeds Aladin Lite~v2 \citep{boch2014}, an interactive sky globe rendered in a \texttt{<canvas>} element within the browser. Survey footprints are loaded as \moc{} overlays using the native Aladin Lite \moc{} rendering pipeline. The user may rotate and zoom the globe interactively. An anchor \moc{} (a small, invisible reference region) is maintained at all times so that the Aladin Lite instance does not collapse when no survey is selected. A colour-coded legend bar appears along the bottom of the globe and updates dynamically as surveys are selected or deselected.

\subsubsection{Equirectangular View}

A second view presents a full-sky equirectangular (plate carr\'{e}e) projection rendered as an SVG element using D3.js~v7 \citep{bostock2011}. Right ascension runs from $360^{\circ}$ to $0^{\circ}$ (left to right, matching the astronomical convention) and declination from $-90^{\circ}$ to $+90^{\circ}$ (bottom to top). Survey footprints in this view are rendered from pre-computed GeoJSON files (one per survey) rather than directly from the FITS \moc{}, avoiding in-browser FITS parsing for this projection. Optional overlays, such as the coordinate grid (RA every $30^{\circ}$, Dec every $30^{\circ}$), the Galactic plane, and the Ecliptic plane, can be toggled independently via controls in the map bottom bar. An in-map legend with coloured rectangular swatches and survey labels is rendered directly on the SVG, providing a publication-ready figure with a single click. The equirectangular view can be exported as a PDF document via the ``Download map (PDF)'' button.

\subsection{Cross-Match Mode}

Activating the ``Show cross-match only'' toggle switches both map views to a cross-match display mode. In this mode, all selected survey overlays are desaturated to a grey colour, and only the region of sky common to all selected surveys is highlighted in white. The intersection area reported in the Coverage panel is updated to reflect the cross-matched region. This mode facilitates rapid visual and quantitative assessment of survey co-coverage.

\subsection{Source Catalogue Upload and Augmentation}

Users may upload a source catalogue in comma-separated (CSV) or tab-separated (TSV) format. The file must contain columns named \texttt{ra} and \texttt{dec} (case-insensitive); additional column name variants such as \texttt{radeg}, \texttt{decl}, and \texttt{declination} are also recognised through fuzzy matching. Right ascension values recorded in hours are automatically detected (either from the column name, e.g., \texttt{ra\_hr}, or from the value range 0--24) and converted to degrees before plotting.

Once uploaded, sources are plotted as pink circles on both the Aladin globe and the equirectangular map. For each selected survey, the tool computes whether each source falls within the survey footprint using the \texttt{filterCoos()} operation of the WebAssembly \moc{} engine (Section~\ref{sec:wasm}). The user can then download an augmented version of the catalogue as a CSV file with a boolean column (true~=~inside, false~=~outside) appended for each selected survey. Column names follow the pattern \texttt{<survey\_id>}, e.g., \texttt{euclid}, \texttt{lsst\_wfd}. An example is demonstrated in Sec.~\ref{sec:survey_cross_match}.

\subsection{Custom \moc{} Upload} 

In addition to the built-in surveys, users may upload their own sky region as a \moc{} in FITS format. The uploaded \moc{} is displayed as an overlay on both map views with a distinct pink colour and a labelled legend entry. The labelled legend entry matches the uploaded filename. It participates fully in intersection calculations and catalogue membership testing, enabling cross-matching against user-defined regions as well as the standard survey footprints. 

\subsection{Colour Themes}

Three perceptual colour palettes are provided, selectable from a dropdown in the sidebar: \textbf{Rainbow} (default), the Paul~Tol 14-colour Discrete Rainbow palette, providing maximum spectral contrast between simultaneously displayed surveys; \textbf{Iridescent}, a sequential blue-to-plum palette from the Paul~Tol Iridescent scheme, suitable for ordered or ranked comparisons; and \textbf{Vivid}, a high-saturation variant optimised for presentations and printed posters. All three palettes are designed to be accessible to readers with colour vision deficiencies \citep{tol2021}.

\subsection{Dark and Light UI Themes}

The application supports both a dark UI theme (default, dark navy background) and a light UI theme (light grey background), toggled by a switch in the topbar.

The equirectangular map background colour adapts to the active UI theme to maintain the sky overlay's legibility in both modes. The preference is persisted to \texttt{localStorage}.

\subsection{Session Persistence}

When the ``Remember selections on this device'' preference is enabled, the full application state is serialised to browser \texttt{localStorage} on every change. This includes the active survey selection and ordering, colour theme, projection, cross-match mode, and any uploaded catalogue data. The state is automatically restored on the next page load, enabling a work session to resume without re-uploading data.

%=============================================================================
\section{Survey Footprints}
\label{sec:surveys}
%=============================================================================

\subsection{Multi-Order Coverage Maps}

Survey footprints in the \sce{} are stored and processed as Multi-Order Coverage (\moc{}) maps, the IVOA standard format for representing sky regions on the sphere
\citep{fernique2014, fernique2022}. \moc{} maps use the HEALPix hierarchical pixelisation of the sphere \citep{gorski2005} at adaptively chosen resolutions, encoding a sky region as a sorted list of HEALPix cell indices. This representation supports exact set-theoretic operations (union, intersection, difference, complement) through simple sorted-list arithmetic, making it well-suited for multi-survey overlap calculations.

All \moc{} FITS files in the \sce{} are stored using NUNIQ pixel encoding with the \texttt{pre\_v2=True} flag to ensure compatibility with Aladin Lite~v2. The HEALPix order used for each survey reflects a compromise between angular resolution and file size; orders in the range 7 to 9 (pixel scales of $\approx$28--7~arcmin) are used throughout.

\subsection{MOC Generation Pipeline}

Survey \moc{} files were generated from publicly available coverage masks, polygon vertices, and pointing lists using standard HEALPix operations. HEALPix pixel indices covered by each survey are identified and converted to MOC format using \texttt{mocpy} \citep{mocpy} with NUNIQ encoding and the \texttt{pre\_v2=True} flag to ensure compatibility with Aladin Lite~v2.

\subsection{Included Surveys}

Table~\ref{tab:surveys} summarises the thirteen survey footprints currently
included in version~2.5.0 of the \sce{}.

\begin{table*}[h!]
\centering
\caption{Survey footprints included in \sce{} v2.5.0. Areas are computed
directly from the distributed FITS \moc{} files using
\texttt{moc.sky\_fraction}~$\times$~41\,252.96~deg$^{2}$. Reported values are MOC-definition dependent; they describe the exact footprint products currently distributed with the application.}
\label{tab:surveys}
\setlength{\tabcolsep}{4pt}
\small
\begin{tabularx}{\linewidth}{lll r >{\raggedright\arraybackslash}X}
\toprule
Survey & Label & Wavelength & Area (deg$^{2}$) & Reference \\
\midrule
Kilo-Degree Survey             & KiDS              & opt     & 891.2         & \citet{dejong2013}    \\
Euclid Data Release 1          & Euclid DR1        & opt/NIR & 2\,108.5      & \citet{euclid2025}    \\
Hyper Suprime-Cam Survey       & HSC               & opt/NIR & 1\,653.4      & \citet{aihara2018}    \\
Dark Energy Survey             & DES               & opt/NIR & 5\,155.0      & \citet{des2016}       \\
UNIONS                         & UNIONS            & opt     & 6\,194.2      & \citet{gwyn2025}     \\
DESI Legacy Imaging DR9        & DESI Legacy DR9   & opt/IR  & 20\,813.1     & \citet{dey2019}       \\
eROSITA All-Sky Survey         & eRASS1            & X-ray   & 21\,524.4     & \citet{merloni2024}   \\
LSST Wide-Fast-Deep            & LSST WFD          & opt     & 17\,719.2     & \citet{ivezic2019}    \\
Roman HLWAS (Wide+Medium)      & Roman HLWAS       & NIR     & 5\,314.0      & \citet{akeson2019}    \\
Roman HLWAS Deep Tier          & Roman HLWAS Deep  & opt/NIR & 35.6          & \citet{hirata2024}    \\
Roman HLTDS                    & Roman HLTDS       & opt/NIR & 28.1          & \citet{romanobservatory2025}      \\
Roman HLTDS Deep               & Roman HLTDS Deep  & opt/NIR & 7.7           & \citet{romanobservatory2025}     \\
ACT Legacy Survey              & ACT Legacy        & mm      & 11\,245.0     & \citet{act2026}       \\
\bottomrule
\end{tabularx}
\end{table*}

\subsubsection{Notes on Individual Surveys} 

\paragraph{KiDS:} The KiDS footprint is derived from the KiDS-N and KiDS-S science-quality binary masks (WCS CAR projection, 1~arcmin pixels), which encode the usable area after removal of bright stars, chip gaps, and artefacts. The resulting \moc{} area of 891~deg$^{2}$ is therefore smaller than the full KiDS survey tile footprint ($\approx$1\,350~deg$^{2}$). 

\paragraph{Euclid DR1:} The Euclid Data Release 1 (DR1) footprint is derived from the DR1 input HEALPix coverage map, with all non-zero pixels converted directly to a \moc{}. This corresponds to an area of $\approx$2\,109~deg$^{2}$.

\paragraph{HSC:} The Subaru Hyper Suprime-Cam Survey footprint is derived from the rectangular field definitions in the LAMBDA footprint file\footnote{Here ``LAMBDA footprint file'' refers to the \texttt{footprint.txt} configuration file of the NASA Legacy Archive for Microwave Background Data Analysis (LAMBDA) Footprint Library, which enumerates survey regions using entries such as polygons and rectangle bounds; see \url{https://lambda.gsfc.nasa.gov/toolbox/footprint/configfile.html}; accessed 2026-05-08.}, combining the Fall~1 Equatorial, Fall~2 Equatorial, Spring Equatorial, and North fields into a single \moc{}. The area of $\approx$1\,653~deg$^{2}$, therefore, reflects these rectangle-based footprint definitions.

\paragraph{DES:} The Dark Energy Survey footprint is derived from the \texttt{DES-polygon} boundary in the LAMBDA footprint file, which is converted directly into a polygon-based \moc{}. 

\paragraph{UNIONS:} The Ultraviolet Near-Infrared Optical Northern Survey footprint is derived from the released celestial HEALPix footprint map in the $ugriz$ bands, with finite non-zero pixels converted directly to a \moc{}. This corresponds to an area of $\approx$6\,194~deg$^{2}$. 

\paragraph{DESI Legacy DR9:} The DESI Legacy DR9 footprint is derived from the current brick-union ingestion workflow. Consequently, values in Table~\ref{tab:surveys} should be interpreted as the areas of the specific footprint products distributed with the application, not as canonical survey-wide areas under all selection criteria.

\paragraph{eRASS1:} The ROSITA All-Sky Survey footprint was generated from the SRG/eROSITA first all-sky survey sky tile catalogue \citep{merloni2024}, covering the Western Galactic hemisphere observed by the German eROSITA consortium during the first all-sky scan.

\paragraph{LSST WFD:} The Legacy Survey of Space and Time Wide Fast Deep (WFD) footprint is derived from the current Rubin scheduler area map, selecting the low-dust and Virgo regions and converting the resulting HEALPix pixels to a \moc{}. The $\approx$17\,719~deg$^{2}$ area should therefore be interpreted as the area of this adopted WFD region definition, rather than a universal LSST footprint under all scheduling assumptions. 

\paragraph{Roman HLWAS (Full and Deep Tiers):}
The Nancy Grace Roman Space Telescope has not yet been launched at the time of writing. The HLWAS footprints are based on the ROTAC-approved tiling plan (\citealt{hirata2024}) from the public repository \texttt{hirata10/RomanHLWASTools}. Two separate \moc{} files are generated: \textbf{Roman HLWAS} covers the Wide and Medium imaging tiers (5\,314.0~deg$^{2}$), and \textbf{Roman HLWAS Deep Tier} covers the two deep
calibration fields at COSMOS and XMM-LSS, totalling 35.6~deg$^{2}$. 

\paragraph{Roman HLTDS (Full and Deep Tiers):}
The Roman High-Latitude Time-Domain Survey (HLTDS) footprints are based on an approximate project-local pointing model derived from the published field centres, tier structure, and pointing counts in the Roman User Documentation \citep{romanobservatory2025}.\footnote{\href{https://roman-docs.stsci.edu/roman-community-defined-surveys/high-latitude-time-domain-survey}{https://roman-docs.stsci.edu/roman-community-defined-surveys/high-latitude-time-domain-survey}; accessed 2026-05-08. Minor technical details remain subject to change in the official HLTDS implementation.} The survey is divided into four tiers: North Wide (38~pointings), North Deep (7~pointings), South Wide (27~pointings), and South Deep (16~pointings). Two \moc{} files are generated: \textbf{Roman HLTDS} combines all four tiers (88~pointings, 28.1~deg$^{2}$), and \textbf{Roman HLTDS Deep} covers only the two deep tiers (23~pointings, 7.7~deg$^{2}$). 

\paragraph{ACT Legacy:} The Atacama Cosmology Telescope Legacy footprint is intended to represent the union of the ACT EQU and South survey regions, following the LAMBDA Footprint Library definitions, which reference the released 148\,GHz HEALPix hit maps for the two components. 

% =============================================================================
\section{Technical Implementation}
\label{sec:implementation}
% =============================================================================

\subsection{Frontend Stack}

The \sce{} is implemented as a static single-page web application requiring no server-side computation. The frontend is written in vanilla JavaScript (ES2022~modules) and depends on four external libraries loaded from CDN: Aladin Lite~v2 \citep{boch2014} for interactive sky globe rendering with native \moc{} overlay support; D3.js~v7 \citep{bostock2011} for SVG-based equirectangular map construction and GeoJSON polygon rendering; jsPDF~2.5 for client-side PDF generation; and jQuery~3.7, required by Aladin Lite~v2 for the Document Object Model (DOM) event handling. No build toolchain, bundler, or transpiler is used. The live application is available at \url{https://www.lammimahad.com/survey-footprint-explorer}. 

\subsection{WebAssembly MOC Engine}
\label{sec:wasm}

All \moc{} operations -- coverage area computation, multi-survey intersection, and per-source membership testing -- are performed client-side using a WebAssembly-compiled Rust \moc{} library \citep{mocwasm}. The engine supports all standard set-theoretic operations (union, intersection, difference) on arbitrary sky regions, enabling sub-second computation of multi-survey overlap areas and per-source membership tests without transmitting data to any external server.

%============================================================================
\section{Illustrative Use Cases}
\label{sec:usecases}

We present three worked examples that illustrate the primary use cases of the \sce{}. 

\subsection{Use Case A: Multi-Survey Footprint Comparison and Overlap Quantification}
\label{sec:use_case_A}

A common preliminary task in multi-survey analyses is to quantify the sky area jointly covered by two or more observing programs. As an illustrative case, we consider the overlap among the Euclid DR1 footprint, the LSST Wide-Fast-Deep (WFD) region, and the full Roman High Latitude Wide Area Survey (HLWAS) footprint. In this workflow, the user selects these three surveys from the survey menu, after which the intersection area is reported in the Coverage panel. The shared region can be isolated by enabling the “Show cross-match only” option, which suppresses the full individual footprints and displays only their common sky area in white on both map projections.

\begin{figure*}[!t]
\centering 
\includegraphics[width=0.8\textwidth] {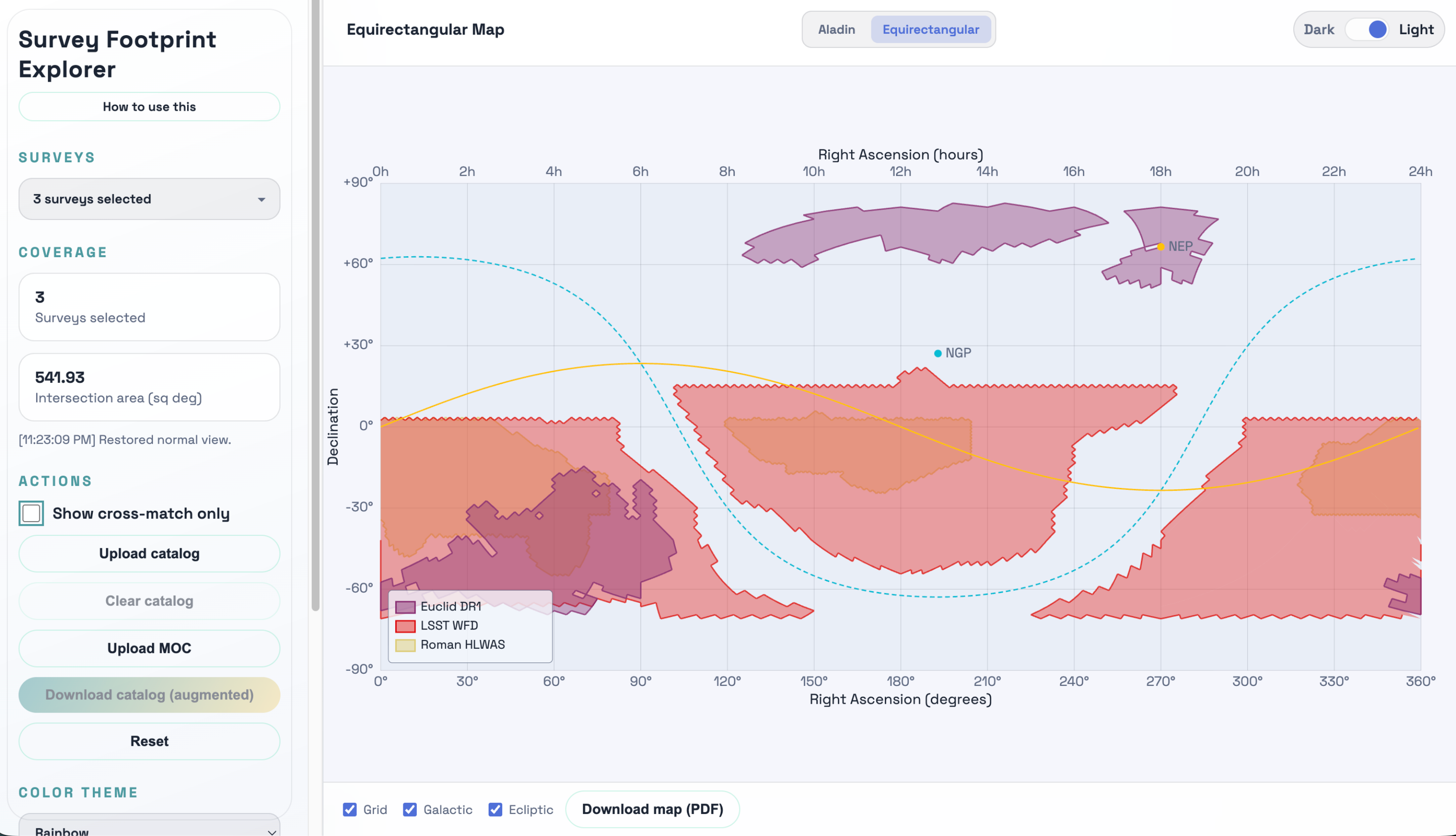} 
\includegraphics[width=0.8\textwidth] {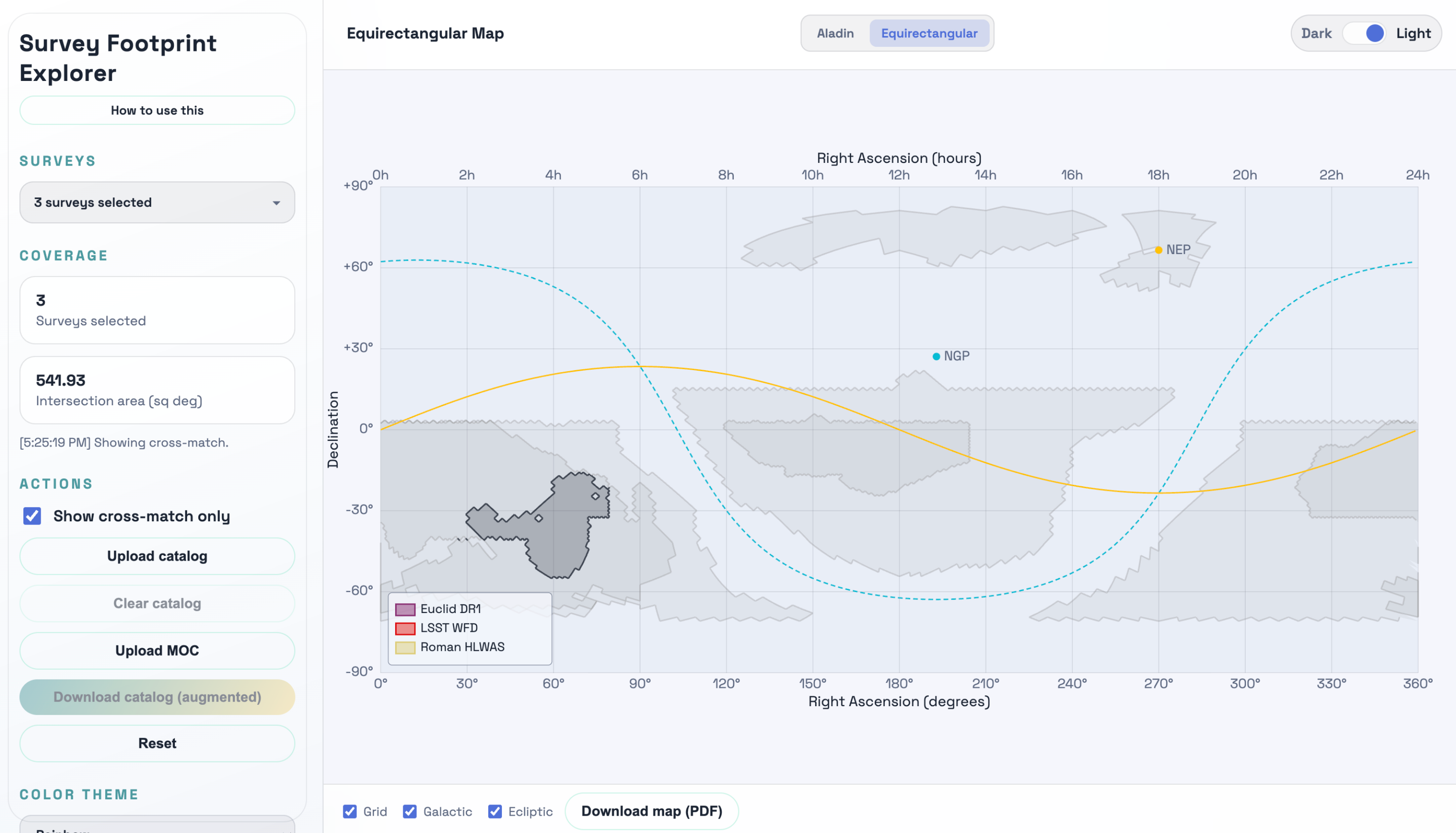} 
\caption{Use Case A: overlap quantification across Euclid~DR1, LSST WFD, and Roman HLWAS. Interface snapshot showing the selected survey workflow. The top panel shows the corresponding survey footprint colours, while the bottom panel shows the overlapping area with ``Show cross-match only'' enabled.}
\label{fig:usecase-overlap}
\end{figure*}

The equirectangular projection is especially useful for interpreting this overlap in an astrophysical context. When the Galactic plane overlay is enabled, regions of the intersection that lie close to the plane can be readily identified, allowing users to assess the extent to which Galactic extinction may contaminate or limit extragalactic applications. A publication-quality visualisation of the resulting overlap can then be exported directly as a PDF using the “Download map (PDF)” control in the lower bar of the equirectangular map (Figure~\ref{fig:usecase-overlap}).

\subsection{Use Case B: Source Catalogue Cross-Matching and Survey Membership}
\label{sec:survey_cross_match}

This functionality is useful for checking whether a sample of target objects (e.g., galaxies, galaxy clusters) is present in one or more surveys. The source catalogue used here is a CSV file with columns \texttt{RA\_deg} and \texttt{DEC\_deg}, and they are checked for the same three surveys as shown in Sec.~\ref{sec:use_case_A}, Euclid DR1, LSST WFD, and Roman HLWAS.

\begin{figure*}[!t]
\centering
\includegraphics[width=0.85\textwidth]{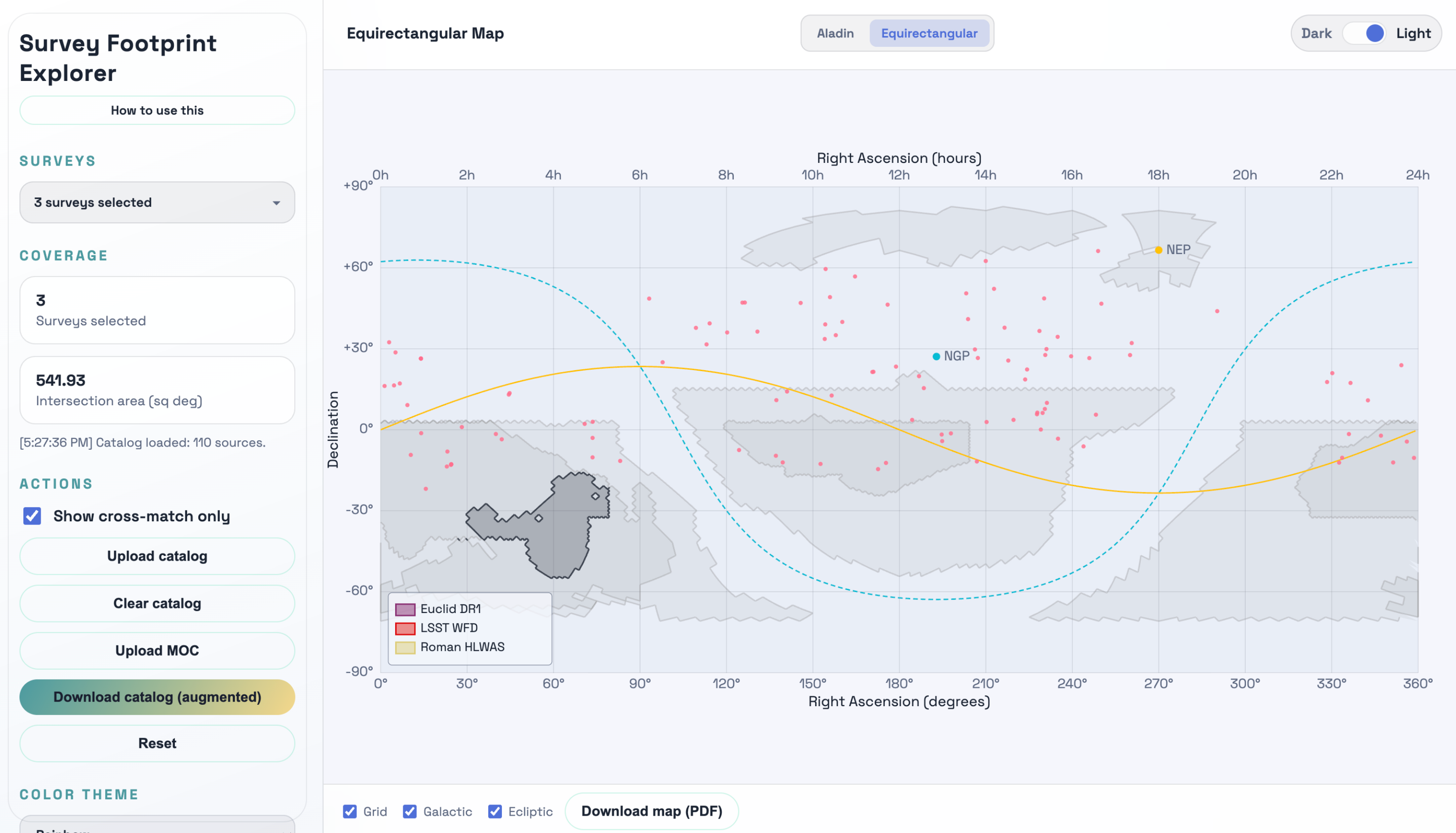}
\caption{Use Case B: source-catalogue membership analysis. App-oriented view
of catalogue upload and survey selection.}
\label{fig:usecase-catalog}
\end{figure*}

The user uploads the catalogue via ``Upload catalog'' in the sidebar. The application detects the \texttt{ra} and \texttt{dec} column names, parses the coordinates, and plots all sources on both map views. The user can select the target surveys before or after the source catalogue is uploaded. After clicking ``Download catalog (augmented)'', the application calls \texttt{filterCoos()} for each survey's \moc{} against the full source list and writes a new CSV file containing three additional boolean columns: \texttt{in\_euclid\_dr1}, \texttt{in\_lsst\_wfd}, and \texttt{in\_roman\_hlwas}. For a catalogue of 50\,000 sources, the membership computation and download can be completed in under two seconds on a typical laptop browser, with all processing occurring locally in the WebAssembly engine; no data is transmitted to any external server (Figure~\ref{fig:usecase-catalog}).

\subsection{Use Case C: Overlay of a Custom Sky Region}

If a user wants to check the overlap of a custom survey footprint, for example, a set of proposed spectroscopic follow-up tiles, as a \moc{} FITS file using \texttt{mocpy}, with any of the listed surveys in the tool, or wants to check a custom source catalogue with the custom survey footprint, they can use this functionality. 

\begin{figure*}[!t]
\centering 
\includegraphics[width=0.85\textwidth] {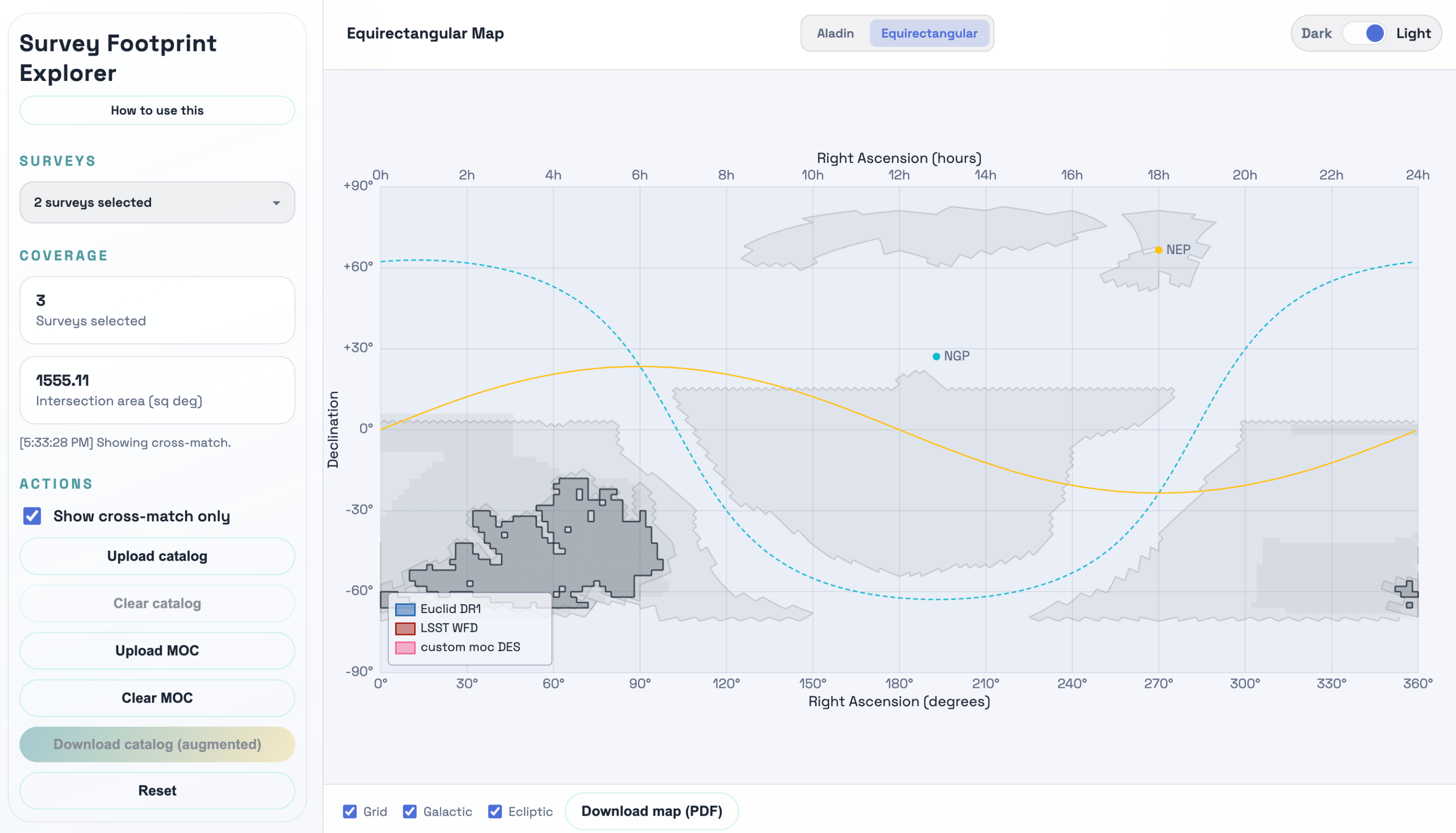} 
\caption{Use Case C: overlap quantification across Euclid~DR1, LSST WFD, and a custom \moc{} region. The interface snapshot is showing the selected survey workflow.}
\label{fig:usecase-custom}
\end{figure*}

The user clicks ``Upload MOC'', selects their FITS file, and the custom region is immediately rendered in both map views with a pink overlay and a labelled legend entry (same as the name of their FITS file). They can select surveys from the survey dropdown menu before or after uploading their \moc{} file to check overlap. The Coverage panel reports the area of intersection between the two regions. Activating ``Show cross-match only'' highlights the sky area covered by both the custom \moc{} and the selected surveys. If a catalogue of tile centres is available, the user can upload it and download the augmented CSV to flag which tiles fall within the target survey footprint, facilitating rapid prioritisation of follow-up resources (Figure~\ref{fig:usecase-custom}).

% =============================================================================
\section{Summary and Future Work}
\label{sec:conclusion}
% =============================================================================

\subsection{Summary}

We presented the \sce{} (version~2.5.0), a browser-based interactive tool for visualising, comparing, and cross-matching the footprints of major astronomical imaging surveys. The key properties of the tool are:

\begin{itemize}
  \item \textbf{Zero-install access:} The application runs entirely in the browser as a static web page, requiring no Python environment, no local software installation, and no server connection for its core functionality.

  \item \textbf{Thirteen built-in survey footprints:} spanning wavelengths from X-ray (eRASS1) to near-infrared (Euclid~DR1, Roman HLWAS), with per-survey coverage areas from 7.7 to 21\,524.4~deg$^{2}$. For the current v2.5.0 MOC set, the sum of individual survey areas is 80\,880.1~deg$^{2}$, while the union sky coverage is 34\,104.7~deg$^{2}$ due to substantial overlap.

  \item \textbf{Client-side \moc{} engine:} All coverage area computations and per-source membership tests are performed using a WebAssembly-compiled Rust \moc{} engine, enabling sub-second intersection calculations for catalogues of tens of thousands of sources with data contained within the user's browser.

  \item \textbf{Dual map views:} An interactive Aladin Lite~v2 globe provides familiar sky navigation, while an SVG-based equirectangular projection offers a publication-ready full-sky view with optional Galactic and Ecliptic overlays and one-click PDF export.

  \item \textbf{Catalogue augmentation pipeline:} Users may upload CSV or TSV catalogues and download an augmented version with per-survey boolean membership columns, facilitating rapid sample selection for multi-survey analyses.
    
  \item \textbf{Custom \moc{} inclusion:} Users may upload custom survey footprint as a \moc{} FITS file, and perform all the other functionalities of the tool that are available for the built-in surveys. The user's custom footprint information, similar to all the other user side data, is only available to the user.
  
\end{itemize}

The tool is freely available at \url{https://www.lammimahad.com/survey-footprint-explorer}. \\
An illustrated user guide is available at \url{https://www.lammimahad.com/survey-explorer-instructions}. \\

\subsection{Future Work}

We will continue to include more publicly available large survey footprints. Any updates will be listed in the illustrated user guide. Any suggestions for including specific surveys are welcome. 

%=============================================================================
\section*{Acknowledgements}
%=============================================================================

The authors thank Matt Hilton for sharing the ACT survey footprints. 
The \sce{} was developed in part with the assistance of Claude Code (Anthropic), an AI coding assistant \citep{anthropic2024}. Claude Code contributed to the implementation of the survey processor Python pipeline, the Roman HLWAS \moc{} generation code, and portions of the JavaScript application logic. This development approach, in which AI tools accelerate implementation while all scientific decisions, validation, and design choices remain with the human authors, follows emerging practice in astronomical software development. In accordance with the authorship guidelines of major journals (Nature Portfolio, AAS journals, A\&A), Claude Code is not listed as an author; its contribution is explicitly disclosed here.

\textbf{Author contributions:}
S.~L.~Ahad: Conceptualization, software, visualization, writing.
R.~Brilenkov: Software, writing.
J.~E.~Taylor: Conceptualization, review.

% =============================================================================
% BIBLIOGRAPHY
% =============================================================================

% \bibliography{paper}
\printbibliography

@article{fernique2014,
  author  = {Fernique, P. and Allen, M. and Boch, T. and Chaitra and
             Donaldson, T. and Durand, D. and Genova, F. and Good, J. and
             Grabowski, M. and Hanisch, R. and Louys, M. and Mann, R. and
             Ochsenbein, F. and Taylor, M. and Wicenec, A.},
  title   = {{MOC --- HEALPix Multi-Order Coverage map}},
  journal = {\aap},
  year    = {2015},
  volume  = {578},
  pages   = {A114},
  doi     = {10.1051/0004-6361/201526075},
  adsurl  = {https://ui.adsabs.harvard.edu/abs/2015A%26A...578A.114F}
}

@misc{fernique2022,
  author       = {Fernique, P. and Boch, T. and Donaldson, T. and
                  Durand, D. and Genova, F. and Louys, M. and
                  Taylor, M. and Wicenec, A.},
  title        = {{Multi-Order Coverage (MOC) Version 2.0}},
  howpublished = {IVOA Recommendation},
  year         = {2022},
  url          = {https://ivoa.net/documents/MOC/}
}

@article{gorski2005,
  author  = {G{\'o}rski, K.~M. and Hivon, E. and Banday, A.~J. and
             Wandelt, B.~D. and Hansen, F.~K. and Reinecke, M. and
             Bartelmann, M.},
  title   = {{HEALPix: A Framework for High-Resolution Discretization and
              Fast Analysis of Data Distributed on the Sphere}},
  journal = {\apj},
  year    = {2005},
  volume  = {622},
  pages   = {759--771},
  doi     = {10.1086/427976},
  adsurl  = {https://ui.adsabs.harvard.edu/abs/2005ApJ...622..759G}
}

@article{zonca2019,
  author  = {Zonca, A. and Singer, L.~P. and Lenz, D. and Reinecke, M. and
             Rosset, C. and Hivon, E. and G{\'o}rski, K.~M.},
  title   = {{healpy: equal area pixelization and spherical harmonics
              transforms for data on the sphere}},
  journal = {Journal of Open Source Software},
  year    = {2019},
  volume  = {4},
  number  = {34},
  pages   = {1298},
  doi     = {10.21105/joss.01298},
  adsurl  = {https://ui.adsabs.harvard.edu/abs/2019JOSS....4.1298Z}
}

@misc{mocpy,
  author       = {Boch, T. and Donaldson, T. and Durand, D. and
                  Fernique, P. and Pineau, F.-X.},
  title        = {{mocpy: MOC Python library}},
  howpublished = {GitHub repository},
  year         = {2023},
  url          = {https://github.com/cds-astro/mocpy}
}

@misc{mocwasm,
  author       = {{CDS Strasbourg}},
  title        = {{cds-moc-rust: Rust implementation of the MOC standard,
                   compiled to WebAssembly}},
  howpublished = {GitHub repository},
  year         = {2023},
  url          = {https://github.com/cds-astro/cds-moc-rust}
}

@inproceedings{boch2014,
  author    = {Boch, T. and Fernique, P.},
  title     = {{Aladin Lite: Embed your Sky Atlas in the Browser}},
  booktitle = {Astronomical Data Analysis Software and Systems XXIII},
  series    = {ASP Conf. Ser.},
  volume    = {485},
  editor    = {Manset, N. and Forshay, P.},
  year      = {2014},
  pages     = {277--280},
  adsurl    = {https://ui.adsabs.harvard.edu/abs/2014ASPC..485..277B}
}

@article{bostock2011,
  author  = {Bostock, M. and Ogievetsky, V. and Heer, J.},
  title   = {{D3: Data-Driven Documents}},
  journal = {IEEE Transactions on Visualization and Computer Graphics},
  year    = {2011},
  volume  = {17},
  number  = {12},
  pages   = {2301--2309},
  doi     = {10.1109/TVCG.2011.185}
}

@inproceedings{taylor2005,
  author    = {Taylor, M.~B.},
  title     = {{TOPCAT \& STIL: Starlink Table Infrastructure Library
                and Applications}},
  booktitle = {Astronomical Data Analysis Software and Systems XIV},
  series    = {ASP Conf. Ser.},
  volume    = {347},
  editor    = {Shopbell, P. and Britton, M. and Ebert, R.},
  year      = {2005},
  pages     = {29--33},
  adsurl    = {https://ui.adsabs.harvard.edu/abs/2005ASPC..347...29T}
}

@misc{tol2021,
  author       = {Tol, P.},
  title        = {{Colour Schemes}},
  howpublished = {SRON Technical Note SRON/EPS/TN/09-002},
  year         = {2021},
  url          = {https://personal.sron.nl/~pault/}
}

@ARTICLE{euclid2025,
       author = {{Euclid Collaboration} and {Mellier}, Y. and {Abdurro'uf} and {Acevedo Barroso}, J.~A. and {Ach{\'u}carro}, A. and {Adamek}, J. and {Adam}, R. and {Addison}, G.~E. and {Aghanim}, N. and {Aguena}, M. and {Ajani}, V. and {Akrami}, Y. and {Al-Bahlawan}, A. and {Alavi}, A. and {Albuquerque}, I.~S. and {Alestas}, G. and {Alguero}, G. and {Allaoui}, A. and {Allen}, S.~W. and {Allevato}, V. and {Alonso-Tetilla}, A.~V. and {Altieri}, B. and {Alvarez-Candal}, A. and {Alvi}, S. and {Amara}, A. and {Amendola}, L. and {Amiaux}, J. and {Andika}, I.~T. and {Andreon}, S. and {Andrews}, A. and {Angora}, G. and {Angulo}, R.~E. and {Annibali}, F. and {Anselmi}, A. and {Anselmi}, S. and {Arcari}, S. and {Archidiacono}, M. and {Aric{\`o}}, G. and {Arnaud}, M. and {Arnouts}, S. and {Asgari}, M. and {Asorey}, J. and {Atayde}, L. and {Atek}, H. and {Atrio-Barandela}, F. and {Aubert}, M. and {Aubourg}, E. and {Auphan}, T. and {Auricchio}, N. and {Aussel}, B. and {Aussel}, H. and {Avelino}, P.~P. and {Avgoustidis}, A. and {Avila}, S. and {Awan}, S. and {Azzollini}, R. and {Baccigalupi}, C. and {Bachelet}, E. and {Bacon}, D. and {Baes}, M. and {Bagley}, M.~B. and {Bahr-Kalus}, B. and {Balaguera-Antolinez}, A. and {Balbinot}, E. and {Balcells}, M. and {Baldi}, M. and {Baldry}, I. and {Balestra}, A. and {Ballardini}, M. and {Ballester}, O. and {Balogh}, M. and {Ba{\~n}ados}, E. and {Barbier}, R. and {Bardelli}, S. and {Baron}, M. and {Barreiro}, T. and {Barrena}, R. and {Barriere}, J.-C. and {Barros}, B.~J. and {Barthelemy}, A. and {Bartolo}, N. and {Basset}, A. and {Battaglia}, P. and {Battisti}, A.~J. and {Baugh}, C.~M. and {Baumont}, L. and {Bazzanini}, L. and {Beaulieu}, J.-P. and {Beckmann}, V. and {Belikov}, A.~N. and {Bel}, J. and {Bellagamba}, F. and {Bella}, M. and {Bellini}, E. and {Benabed}, K. and {Bender}, R. and {Benevento}, G. and {Bennett}, C.~L. and {Benson}, K. and {Bergamini}, P. and {Bermejo-Climent}, J.~R. and {Bernardeau}, F. and {Bertacca}, D. and {Berthe}, M. and {Berthier}, J. and {Bethermin}, M. and {Beutler}, F. and {Bevillon}, C. and {Bhargava}, S. and {Bhatawdekar}, R. and {Bianchi}, D. and {Bisigello}, L. and {Biviano}, A. and {Blake}, R.~P. and {Blanchard}, A. and {Blazek}, J. and {Blot}, L. and {Bosco}, A. and {Bodendorf}, C. and {Boenke}, T. and {B{\"o}hringer}, H. and {Boldrini}, P. and {Bolzonella}, M. and {Bonchi}, A. and {Bonici}, M. and {Bonino}, D. and {Bonino}, L. and {Bonvin}, C. and {Bon}, W. and {Booth}, J.~T. and {Borgani}, S. and {Borlaff}, A.~S. and {Borsato}, E. and {Bose}, B. and {Botticella}, M.~T. and {Boucaud}, A. and {Bouche}, F. and {Boucher}, J.~S. and {Boutigny}, D. and {Bouvard}, T. and {Bouwens}, R. and {Bouy}, H. and {Bowler}, R.~A.~A. and {Bozza}, V. and {Bozzo}, E. and {Branchini}, E. and {Brando}, G. and {Brau-Nogue}, S. and {Brekke}, P. and {Bremer}, M.~N. and {Brescia}, M. and {Breton}, M.-A. and {Brinchmann}, J. and {Brinckmann}, T. and {Brockley-Blatt}, C. and {Brodwin}, M. and {Brouard}, L. and {Brown}, M.~L. and {Bruton}, S. and {Bucko}, J. and {Buddelmeijer}, H. and {Buenadicha}, G. and {Buitrago}, F. and {Burger}, P. and {Burigana}, C. and {Busillo}, V. and {Busonero}, D. and {Cabanac}, R. and {Cabayol-Garcia}, L. and {Cagliari}, M.~S. and {Caillat}, A. and {Caillat}, L. and {Calabrese}, M. and {Calabro}, A. and {Calderone}, G. and {Calura}, F. and {Camacho Quevedo}, B. and {Camera}, S. and {Campos}, L. and {Ca{\~n}as-Herrera}, G. and {Candini}, G.~P. and {Cantiello}, M. and {Capobianco}, V. and {Cappellaro}, E. and {Cappelluti}, N. and {Cappi}, A. and {Caputi}, K.~I. and {Cara}, C. and {Carbone}, C. and {Cardone}, V.~F. and {Carella}, E. and {Carlberg}, R.~G. and {Carle}, M. and {Carminati}, L. and {Caro}, F. and {Carrasco}, J.~M. and {Carretero}, J. and {Carrilho}, P. and {Carron Duque}, J. and {Carry}, B.},
        title = "{Euclid: I. Overview of the Euclid mission}",
      journal = {A\&A},
         year = 2025,
        month = may,
       volume = {697},
          eid = {A1},
        pages = {A1},
          doi = {10.1051/0004-6361/202450810},
archivePrefix = {arXiv},
       eprint = {2405.13491},
 primaryClass = {astro-ph.CO},
       adsurl = {https://ui.adsabs.harvard.edu/abs/2025A&A...697A...1E}
}

@article{ivezic2019,
  author  = {Ivezi{\'c}, {\v Z}. and Kahn, S.~M. and Tyson, J.~A. and
             Abel, B. and Acosta, E. and Allsman, R. and others},
  title   = {{LSST: From Science Drivers to Reference Design and
              Anticipated Data Products}},
  journal = {\apj},
  year    = {2019},
  volume  = {873},
  pages   = {111},
  doi     = {10.3847/1538-4357/ab042c},
  adsurl  = {https://ui.adsabs.harvard.edu/abs/2019ApJ...873..111I}
}

@misc{akeson2019,
  author       = {Akeson, R. and Armus, L. and Bachelet, E. and Bailey, V. and
                  Bartusek, L. and Bellini, A. and others},
  title        = {{The Wide Field Infrared Survey Telescope: 100 Hubbles
                   for the 2020s}},
  howpublished = {arXiv:1902.05569},
  year         = {2019},
  url          = {https://arxiv.org/abs/1902.05569},
  adsurl       = {https://ui.adsabs.harvard.edu/abs/2019arXiv190205569A}
}

@misc{hirata2024,
  author       = {Hirata, C.},
  title        = {{RomanHLWASTools: ROTAC-approved tiling data for the
                   Roman HLWAS}},
  howpublished = {GitHub repository},
  year         = {2024},
  url          = {https://github.com/hirata10/RomanHLWASTools}
}

@article{dey2019,
  author  = {Dey, A. and Schlegel, D.~J. and Lang, D. and Blum, R. and
             Burleigh, K. and Fan, X. and others},
  title   = {{Overview of the DESI Legacy Imaging Surveys}},
  journal = {\aj},
  year    = {2019},
  volume  = {157},
  pages   = {168},
  doi     = {10.3847/1538-3881/ab089d},
  adsurl  = {https://ui.adsabs.harvard.edu/abs/2019AJ....157..168D}
}

@article{des2016,
  author  = {{Dark Energy Survey Collaboration} and Abbott, T. and
             Abdalla, F.~B. and Aleksi{\'c}, J. and Allam, S. and
             Amara, A. and others},
  title   = {{The Dark Energy Survey: more than dark energy --- an overview}},
  journal = {\mnras},
  year    = {2016},
  volume  = {460},
  pages   = {1270--1299},
  doi     = {10.1093/mnras/stw641},
  adsurl  = {https://ui.adsabs.harvard.edu/abs/2016MNRAS.460.1270D}
}

@article{aihara2018,
  author  = {Aihara, H. and Arimoto, N. and Armstrong, R. and
             Arnouts, S. and Bahcall, N.~A. and Bickerton, S. and others},
  title   = {{The Hyper Suprime-Cam SSP Survey: Overview and Survey Design}},
  journal = {PASJ},
  year    = {2018},
  volume  = {70},
  pages   = {S4},
  doi     = {10.1093/pasj/psx066},
  adsurl  = {https://ui.adsabs.harvard.edu/abs/2018PASJ...70S...4A}
}

@article{dejong2013,
  author  = {de Jong, J.~T.~A. and Verdoes Kleijn, G.~A. and
             Kuijken, K.~H. and Valentijn, E.~A.},
  title   = {{The Kilo-Degree Survey}},
  journal = {Experimental Astronomy},
  year    = {2013},
  volume  = {35},
  pages   = {25--44},
  doi     = {10.1007/s10686-012-9306-1},
  adsurl  = {https://ui.adsabs.harvard.edu/abs/2013ExA....35...25D}
}

@article{merloni2024,
  author  = {Merloni, A. and Lamer, G. and Liu, T. and Ramos-Ceja, M.~E. and
             Brunner, H. and Bulbul, E. and others},
  title   = {{The SRG/eROSITA all-sky survey. First X-ray maps and cluster
              catalogue: first all-sky survey (eRASS1)}},
  journal = {\aap},
  year    = {2024},
  volume  = {682},
  pages   = {A34},
  doi     = {10.1051/0004-6361/202347165},
  adsurl  = {https://ui.adsabs.harvard.edu/abs/2024A%26A...682A..34M}
}

@misc{anthropic2024,
  author       = {{Anthropic}},
  title        = {{Claude Code: AI-powered coding assistant}},
  howpublished = {Software},
  year         = {2024},
  url          = {https://claude.ai/claude-code}
}

@ARTICLE{gwyn2025,
       author = {{Gwyn}, Stephen and {McConnachie}, Alan W. and {Cuillandre}, Jean-Charles and {Chambers}, Ken C. and {Magnier}, Eugene A. and {Hudson}, Michael J. and {Oguri}, Masamune and {Furusawa}, Hisanori and {Hildebrandt}, Hendrik and {Carlberg}, Raymond and {Ellison}, Sara L. and {Furusawa}, Junko and {Gavazzi}, Rapha{\"e}l and {Ibata}, Rodrigo and {Mellier}, Yannick and {Osato}, Ken and {Aussel}, H. and {Baumont}, Lucie and {Bayer}, Manuel and {Boulade}, Olivier and {C{\^o}t{\'e}}, Patrick and {Chemaly}, David and {Daley}, Cail and {Duc}, Pierre-Alain and {Ellien}, A. and {Fabbro}, S{\'e}bastien and {Ferreira}, Leonardo and {Fitriana}, Itsna K. and {Le Floc'h}, Emeric and {Hammer} and {Francois} and {Fudamoto}, Yoshinobu and {Gao}, Hua and {Goh}, L.~W.~K. and {Goto}, Tomotsugu and {Guerrini}, Sacha and {Guinot}, Axel and {H{\'e}nault-Brunet}, Vincent and {Harikane}, Yuichi and {Hayashi}, Kohei and {Heesters}, Nick and {Ichikawa}, Kohei and {Kilbinger}, Martin and {Kuzma}, P.~B. and {Li}, Qinxun and {Liaudat}, Tob{\'\i}as I. and {Lin}, Chien-Cheng and {M{\"u}ller}, Oliver and {Martin}, Nicolas F. and {Matsuoka}, Yoshiki and {Medina}, Gustavo E. and {Miyatake}, Hironao and {Miyazaki}, Satoshi and {Mpetha}, Charlie T. and {Nagao}, Tohru and {Navarro}, Julio F. and {Niwano}, Masafumi and {Ogami}, Itsuki and {Okabe}, Nobuhiro and {Onoue}, Masafusa and {Paek}, Gregory S.~H. and {Parker}, Laura C. and {Patton}, David R. and {Hervas Peters}, Fabian and {Prunet}, Simon and {S{\'a}nchez-Janssen}, Rub{\'e}n and {Schultheis}, M. and {Sestito}, Federico and {Smith}, Simon E.~T. and {Starck}, J. -L. and {Starkenburg}, Else and {Stone}, Connor and {Storfer}, Christopher and {Suzuki}, Yoshihisa and {Erben} and {T.} and {Taibi}, Salvatore and {Thomas}, G.~F. and {TianFang}, Zhang and {Toba}, Yoshiki and {Uchiyama}, Hisakazu and {Valls-Gabaud}, David and {Venn}, Kim A. and {Van Waerbeke}, Ludovic and {Wainscoat}, Richard J. and {Wilkinson}, Scott and {Wittje}, Anna and {Yoshida}, Taketo and {Zhong}, Yuxing},
        title = "{UNIONS: The Ultraviolet Near-Infrared Optical Northern Survey}",
      journal = {arXiv e-prints},
         year = 2025,
        month = mar,
          eid = {arXiv:2503.13783},
        pages = {arXiv:2503.13783},
          doi = {10.48550/arXiv.2503.13783},
archivePrefix = {arXiv},
       eprint = {2503.13783},
 primaryClass = {astro-ph.GA},
       adsurl = {https://ui.adsabs.harvard.edu/abs/2025arXiv250313783G}
}

@ARTICLE{act2026,
       author = {{Aguena}, M. and {Aiola}, S. and {Allam}, S. and {Andrade-Oliveira}, F. and {Bacon}, D. and {Bahcall}, N. and {Battaglia}, N. and {Battistelli}, E.~S. and {Bocquet}, S. and {Bolliet}, B. and {Bond}, J.~R. and {Brooks}, D. and {Calabrese}, E. and {Carretero}, J. and {Choi}, S.~K. and {da Costa}, L.~N. and {Costanzi}, M. and {Coulton}, W. and {Davis}, T.~M. and {Desai}, S. and {Devlin}, M.~J. and {Dicker}, S. and {Doel}, P. and {Duivenvoorden}, A.~J. and {Dunkley}, J. and {Ferraro}, S. and {Flaugher}, B. and {Frieman}, J. and {Gallardo}, P.~A. and {Gatti}, M. and {Gaztanaga}, E. and {Gill}, A.~S. and {Golec}, J.~E. and {Gruen}, D. and {Gruendl}, R.~A. and {Halpern}, M. and {Hasselfield}, M. and {Hill}, J.~C. and {Hilton}, M. and {Hincks}, A.~D. and {Hinton}, S.~R. and {Hollowood}, D.~L. and {Honscheid}, K. and {Hubmayr}, J. and {Huffenberger}, K.~M. and {Hughes}, J.~P. and {James}, D.~J. and {Klein}, M. and {Knowles}, K. and {Koopman}, B.~J. and {Kosowsky}, A. and {Lahav}, O. and {Lee}, E. and {Lin}, Y. and {Lokken}, M. and {Madhavacheril}, M.~S. and {Malag{\'o}n}, A.~A. Plazas and {Marrewijk}, J. v. and {Marshall}, J.~L. and {McMahon}, J. and {Mena-Fern{\'a}ndez}, J. and {Miquel}, R. and {Miyatake}, H. and {Mohr}, J.~J. and {Moodley}, K. and {Mroczkowski}, T. and {Naess}, S. and {Nati}, F. and {Nicola}, A. and {Niemack}, M.~D. and {Ogando}, R.~L.~C. and {Oguri}, M. and {Orlowski-Scherer}, J. and {Page}, L.~A. and {Partridge}, B. and {da Silva Pereira}, M.~E. and {Porredon}, A. and {Qu}, F.~J. and {Ragavan}, D.~C. and {Guachalla}, B. Ried and {Romer}, A.~K. and {Rosell}, A. Carnero and {Rykoff}, E.~S. and {Samuroff}, S. and {Sanchez}, E. and {Sevilla-Noarbe}, I. and {Sierra}, C. and {Sif{\'o}n}, C. and {Smith}, M. and {Staggs}, S.~T. and {Suchyta}, E. and {Swanson}, M.~E.~C. and {Tucker}, D.~L. and {Vargas}, C. and {Vavagiakis}, E.~M. and {De Vicente}, J. and {Weaverdyck}, N. and {Weller}, J. and {Wollack}, E.~J. and {Zubeldia}, I.},
        title = "{The Atacama Cosmology Telescope: DR6 Sunyaev-Zel'dovich Selected Galaxy Clusters Catalog}",
      journal = {The Open Journal of Astrophysics},
         year = 2026,
        month = jan,
       volume = {9},
        pages = {55863},
          doi = {10.33232/001c.155863},
archivePrefix = {arXiv},
       eprint = {2507.21459},
 primaryClass = {astro-ph.CO},
       adsurl = {https://ui.adsabs.harvard.edu/abs/2026OJAp....955863A}
}

@ARTICLE{romanobservatory2025,
       author = {{Observations Time Allocation Committee}, Roman and {Community Survey Definition Committees}, Core},
        title = "{Roman Observations Time Allocation Committee: Final Report and Recommendations}",
      journal = {arXiv e-prints},
         year = 2025,
        month = may,
          eid = {arXiv:2505.10574},
        pages = {arXiv:2505.10574},
          doi = {10.48550/arXiv.2505.10574},
archivePrefix = {arXiv},
       eprint = {2505.10574},
 primaryClass = {astro-ph.IM},
       adsurl = {https://ui.adsabs.harvard.edu/abs/2025arXiv250510574O}
}

% \end{multicols}
\end{document}